\documentclass[twocolumn,showpacs,preprintnumbers,amsmath,amssymb,prb,footinbib]{revtex4-1}

\usepackage{graphicx}
\usepackage[abs]{overpic}
\usepackage{color}

\begin{document}

\title{Charge and spin pumping through a double quantum dot}
\author{Roman-Pascal~Riwar}
\author{Janine~Splettstoesser}
\affiliation{Institut f\"{u}r Theorie der statistischen Physik, RWTH Aachen University, D-52056 Aachen, Germany}
\affiliation{JARA - Fundamentals of Future Information Technology}
\date{\today}

\begin{abstract}
We calculate adiabatic charge and spin pumping through a serial double quantum dot  with strong Coulomb interaction, coupled  to normal metal or ferromagnetic contacts. We use a real-time diagrammatic approach in the regime of weak coupling to the reservoirs. In the case of weak interdot tunnel coupling we investigate the influence of  tunnel-induced renormalization effects due to charge fluctuations on the pumped charge and spin. We show that tunneling through thermally excited states can play an important role in the strong interdot coupling regime. In particular, for ferromagnetic contacts, both effects enable the generation of pure spin currents. Furthermore they can lead to an inverted spin-valve effect or  even to the inversion of the transport direction going along with a diverging tunneling magnetoresistance.
\end{abstract}

 
\maketitle
\section{Introduction}
Pumping through mesoscopic devices is realized in the absence of an external bias by the periodic variation of certain system parameters in time, thus creating directed transport of electrons from one contact to the other. When this variation is slow compared to the characteristic lifetime of the electronic states on the device, one speaks of adiabatic pumping.~\cite{Buttiker94,Brouwer98,Zhou99,Makhlin01,Moskalets01,*Moskalets02a,Entin02} 
In recent years, experiments on quantized pumping were performed, aiming at the realization of a quantum standard for the current or for the repeated initialization of coherent quantum states, useful for quantum operations.~\cite{Pothier92,Fletcher03,*Ebbecke05,*Leek05,Feve07}
Fewer experimental realizations dealt with the role of quantum interference effects giving rise to a directed current due to time-dependent periodic fields.~\cite{Switkes99,Watson03}
Furthermore, when the system through which the charge is pumped is small, as say, quantum dots,  the Coulomb interaction cannot be neglected because of the small capacitance of the system. This impact of electron-electron interactions on adiabatic pumping through quantum dots, quantum wires and metallic islands has lately been considered in theory.~\cite{Citro03,Aono04,Splettstoesser05,Brouwer05,Sela06,Arrachea08,Fioretto08,Hernandez08,Cavaliere09,Winkler09} Particularly, charge pumping can be uniquely due to Coulomb interactions in the non-linear regime~\cite{Reckermann10} or due to interaction-induced renormalization effects.~\cite{Splettstoesser06}\\
Adiabatic pumping requires the time-dependent modulation of at least two of the system's parameters. 
In the work presented here we are interested in charge and spin pumping through a serially coupled double dot, where the energy levels of both quantum dots can be independently varied by time-dependent gate voltages. This was so far examined for noninteracting systems~\cite{Kashcheyevs08,Romeo09} with normal conducting leads, and further studies were performed in the high-frequency regime.~\cite{Hazelzet01,Cota05,*Cota05Err,* Sanchez06,Khomitsky09}
The study of double dots is particularly interesting due to their complex internal spectrum: the eigenstates of the double dot are differently coupled to the leads, and these couplings are effectively energy-level dependent. This results in new effects for pumping, as demonstrated in this paper, which are controllable through the mere modulation of the bare dot energy spectrum. Importantly here, this choice of two independently tunable parameters, i.e., the spectra of the two dots, has an easier experimental access than any available choice of parameters in a single dot. The fact that adiabatic pumping acts as a useful spectroscopy tool,~\cite{Reckermann10} revealing system specific features which are not accessible through a static measurement, establishes one of our motivations to study adiabatic pumping through interacting double dots in various regimes and setups.\\
Previous studies on transport through a double-dot system with Coulomb interaction in the static regime revealed charge fluctuation effects, related to the interplay of the double dot's internal structure and the reservoirs,~\cite{Wunsch05,Konig05} and double dots were studied in an interferometer setup,~\cite{Urban09} to name a few examples. Lately double dots have been extensively proposed and used as measurement devices: the read-out of spin properties is discussed in Ref.~\onlinecite{Hanson07} and references therein, the Pauli spin blockade was used to investigate relaxation times~\cite{Vorontsov08,*Buitelaar08b} where the influence of spin-orbit coupling and the role of nuclear spins have been treated. Furthermore, a double quantum dot device can act as a noise detector.~\cite{Aguado00} In Refs.~\onlinecite{vanderWiel01,*Fujisawa98} transitions in the double-dot spectrum due to a bosonic environment were studied. \\
In addition to the charge degree of freedom also the spin degree of freedom of the electrons plays an important role in transport, in particular, in the context of spintronics~\cite{Prinz98,*Wolf01} and applications in quantum computation.~\cite{Loss98}
The spin-valve effect~\cite{Julliere75,Slonczewski89,Baibich88,*Binasch89,*Moodera96} in systems containing differently polarized ferromagnets and its tunability, e.g., by a gate control~\cite{Datta90,*Schapers00} are therefore of interest.
In the study of \textit{quantum dot} spin valves~\cite{Braun04,Rudzinski01,*Usaj01,*Usaj05,*Konig03,*Choi04,Sahoo05,Cottet06,Trocha09} the effect of Coulomb interaction and the properties of the dot spectrum were taken into account.\\ 
In the field of pumping, spin-dependent charge transport  and spin transport have been studied extensively in the high-frequency regime.~\cite{Cota05,*Cota05Err,* Sanchez06,Wang03,*Fransson10} Adiabatic spin pumping has been explored in both the presence of spin-orbit coupling,~\cite{Governale03,*Brosco09} and finite magnetic fields,~\cite{Mucciolo02,*Blaauboer05} as well as for a quantum dot attached to ferromagnetic contacts.~\cite{Splettstoesser08a}
Recently adiabatic spin pumping in a magnetic wire with domain walls was investigated.~\cite{Zhu10} Adiabatic pumping through double dots in the presence of spin-orbit coupling can be used to study the spin dynamics in a Pauli-blockade configuration.~\cite{Romeo09}
Also charge and spin in ferromagnetic hybrid structures pumped by magnetization dynamics have been investigated in detail.~\cite{Tserkovnyak02,*Brataas02,*Heinrich03,*Xiao08} 
An experimental realization of a spin pump used the modulation of a single quantum dot in the presence of a Zeeman field.~\cite{Watson03}\\
In this paper we study a double-dot system with onsite Coulomb interaction contacted weakly to electronic reservoirs. We consider the situations where both leads are normal conducting (N-DD-N), see Fig.~\ref{fig_Hgeneral}, where only one of the reservoirs is normal conducting while the other is replaced by a ferromagnetic contact (N-DD-F), and finally the case where both contacts are ferromagnetic (F-DD-F).
We investigate the effects of quantum charge fluctuations and the impact of different effective coupling to the hybridized double-dot states on the pumped charge. Based on these effects the pumped charge shows characteristic sign changes. \\
Also in the presence of spin-polarized leads, these two effects matter. With respect to an earlier work on pumping through a single interacting quantum dot,~\cite{Splettstoesser08a} the present study of a double dot reveals a number of strikingly different effects, such as pure spin currents and the inversion of the transport direction due to polarized leads only. \\
We use a real-time diagrammatic approach,~\cite{Konig96a,*Konig96b} extended to the adiabatic regime,~\cite{Splettstoesser06} to calculate the pumped charge and spin, taking into account Coulomb interactions. For the interdot hopping we treat both the case of weak and strong couplings.\\
This paper is organized as follows: we introduce the model and the formalism used for our theoretical investigations in Sec.~\ref{sec_model_formalism}. In the following we present our results for charge pumping through a double dot coupled to normal conductor leads (Sec.~\ref{sec_N_dd_N}). We discuss spin pumping when one of the leads is spin polarized in Sec.~\ref{sec_spin_pump}. In the last part we consider the transport behavior in the presence of two ferromagnetic leads with arbitrary polarization angle (Sec.~\ref{sec_ferro}). We set $\hbar=c=1$ for the rest of this paper.

\section{Model and Formalism}\label{sec_model_formalism}

\subsection{Hamiltonian}\label{sec_hamiltonian}

\begin{figure}[t]
\centering
\includegraphics[scale=1]{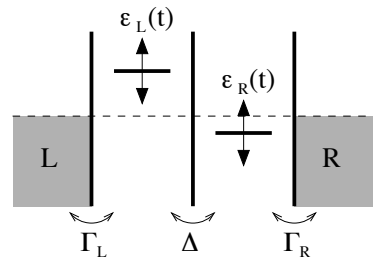}
\caption{Schematic picture of the system. The left and right dot are coupled to the electron reservoirs $\text{L}$ and $\text{R}$, with equal electrochemical potentials, by barriers with tunnel coupling strength $\Gamma_{\text{L}}$ and $\Gamma_{\text{R}}$, respectively. The coupling between the single-level quantum dots is given by $\Delta$ and the two dots have time-dependent energy levels $\epsilon_{\text{L}}\left(t\right)$ and $\epsilon_{\text{R}}\left(t\right)$.}
\label{fig_Hgeneral}
\end{figure}
We consider two single-level spin-degenerate quantum dots coupled in series to each other and tunnel coupled to a left (L) and a right (R) lead, as depicted in Fig.~\ref{fig_Hgeneral}. The Hamiltonian of this system is given by
\begin{equation}
H=H_{\text{dd}}\left(t\right)+\sum_{\alpha=\text{L},\text{R}}H_{\alpha}+H_{\text{tunnel}}\ .
\end{equation}
The Hamiltonian of the double dot is explicitly time dependent,
\begin{equation}
\begin{split}
H_{\text{dd}}\left(t\right)=&\sum_{\alpha=\text{L},\text{R}}\epsilon_{\alpha}\left(t\right)n_{\alpha}+Un_{\text{L}}n_{\text{R}}+U'\left(n_{\text{L}\uparrow}n_{\text{L}\downarrow}+n_{\text{R}\uparrow}n_{\text{R}\downarrow}\right) \\ &-\frac{\Delta}{2}\sum_{\sigma=\uparrow,\downarrow}\left(d_{\text{L}\sigma}^{\dagger}d_{\text{R}\sigma}+d_{\text{R}\sigma}^{\dagger}d_{\text{L}\sigma}\right).
\end{split}
\end{equation}
The single level $\epsilon_{\alpha}\left(t\right)$ of each dot $\alpha=\mathrm{L,R}$ is modulated by periodically time-dependent gate voltages, with a frequency $\Omega$ and a respective phase shift. We define the number operator for spin $\sigma=\uparrow,\downarrow$ on dot $\alpha$ as $n_{\alpha\sigma}=d_{\alpha\sigma}^{\dagger}d_{\alpha\sigma}$ and the total number operator of dot $\alpha$ as $n_{\alpha}=\sum_{\sigma}n_{\alpha\sigma}$, where $d^{\dagger}_{\alpha\sigma}$ ($d_{\alpha\sigma}$) creates (annihilates) an electron with  spin $\sigma$ on the dot $\alpha$. Electrons on each of the dots are subject to on-site Coulomb interaction $U'$ and to Coulomb interaction with electrons on the neighboring dot $U$. The charging energies are related to the capacitances of the dots and are taken into account within a constant interaction model, see, for example, Ref.~\onlinecite{Schon97}.
Hopping from one dot to the other occurs with the interdot coupling amplitude $-\Delta/2$, where $\Delta$ is taken to be real and positive. It is useful to define the mean double-dot energy $E=E\left(t\right)=\left[\epsilon_{\text{L}}\left(t\right)+\epsilon_{\text{R}}\left(t\right)\right]/2=:\bar{E}+\delta E(t)$ as well as the level difference $\epsilon=\epsilon\left(t\right)=\epsilon_{\text{L}}\left(t\right)-\epsilon_{\text{R}}\left(t\right)=:\bar{\epsilon}+\delta \epsilon(t)$, separating each quantity into a time-averaged and a time-dependent part. We will later apply the convention of omitting the time argument and denoting time-averaged quantities by a bar also for functions containing these parameters. In the following we take $U'$ to be the largest energy scale, and thus every dot is at most singly occupied, excluding exchange coupling.

The non-interacting electrons in the left and right leads are described by
\begin{equation}
H_{\alpha}=\sum_{k\sigma}\epsilon_k c^{\dagger}_{\alpha k\sigma}c_{\alpha k\sigma}\ ,
\end{equation}  
with the creation (annihilation) operators $c_{\alpha k\sigma}^{\dagger}$ ($c_{\alpha k\sigma}$) of an electron with spin $\sigma$ and momentum $k$ in lead $\alpha$. We will include both cases of either ferromagnetic or normal conductor leads. We assume the leads to have equal electrochemical potentials, $\mu_{\text{L}}=\mu_{\text{R}}$, not influenced by the time-dependent gate voltages. The effect of a time-dependent bias is discussed in Ref.~\onlinecite{Reckermann10}.
The tunnel Hamiltonian describing the dot-lead coupling is given as
\begin{equation}\label{eq_tunnel_ham}
H_{\text{tunnel}}=\sum_{\alpha=\text{L},\text{R}}\sum_{k,\sigma}V_{\alpha}\left(c_{\alpha k\sigma}^{\dagger}d_{\alpha\sigma}+\mathrm{H.c.}\right)\ ,
\end{equation}
with the tunnel matrix elements $V_{\text{L}}$ and $V_{\text{R}}$, which do not depend on momentum or spin. The tunnel rates are defined as $\Gamma_{\alpha}=2\pi\left|V_{\alpha}\right|^2\rho_{\alpha}$ for $\alpha=\text{L},\text{R}$. We define $\Gamma$ as the sum of the left and right tunnel rates, $\Gamma=\Gamma_\mathrm{L}+\Gamma_\mathrm{R}$. We assume all bare tunneling rates $\Gamma,\Delta$ to be time independent. We will see later on that the tunneling rates to hybrid dot states become effectively time-dependent via the time-dependent dot energy levels. Both reservoirs are taken in the wideband limit where their densities of states $\rho_{\alpha}$ are energy independent. We will later consider ferromagnetic leads by introducing spin-dependent densities of states in the respective lead $\alpha=\text{L},\text{R}$. The technical details for the treatment of ferromagnetic leads will be discussed in Sec.~\ref{sec_ferro}.

\subsection{Real-time diagrammatic approach}\label{sec_diagram}

We describe the system by the reduced density matrix of the double dot; the degrees of freedom of the leads, taking up the role of baths, are traced out. We write the accessible states on the double dot as $\left|\chi\right\rangle$. The time evolution of the density matrix $P_{\chi_1}^{\chi_2}=\langle\chi_2|\rho_\mathrm{dd}|\chi_1\rangle$ is captured in the generalized master equation (kinetic equation)
\begin{equation}\label{eq_master_general}
\begin{split}
\frac{d}{d t}P_{\chi_1}^{\chi_2}\left(t\right)=&-i\sum_{\chi_1',\chi_2'}L_{\chi_1,\chi_1'}^{\chi_2,\chi_2'}\left(t\right)P_{\chi_1'}^{\chi_2'}\left(t\right)+\\ &\sum_{\chi_1',\chi_2'}\int_{-\infty}^{t}dt'W_{\chi_1,\chi_1'}^{\chi_2,\chi_2'}\left(t,t'\right)P_{\chi_1'}^{\chi_2'}\left(t'\right)\ .
\end{split}
\end{equation}
The first part represents the interdot dynamics via a Liouville superoperator, where we define the matrix elements $\langle\chi_2|L(t)\rho_\mathrm{dd}(t)|\chi_1\rangle=\langle\chi_2|[H_\mathrm{dd}(t),\rho_\mathrm{dd}(t)]|\chi_1\rangle=\sum_{\chi_1',\chi_2'}L_{\chi_1,\chi_1'}^{\chi_2,\chi_2'}\left(t\right)P_{\chi_1'}^{\chi_2'}\left(t\right)$. The kernel $W_{\chi_1,\chi_1'}^{\chi_2,\chi_2'}$ in the second part of Eq.~(\ref{eq_master_general}) incorporates the lead-dot tunneling transitions between states $\chi_1'$ and $\chi_2'$ at time $t'$, and states $\chi_1$ and $\chi_2$ at time $t$ and is calculated using a real-time diagrammatic approach.~\cite{Konig96a,*Konig96b}  We write the diagonal and off-diagonal elements of the density matrix in vector form, $\boldsymbol{P}=\left(\{P_{\chi}^{\chi'}\}_{\chi=\chi'},\{P_{\chi}^{\chi'}\}_{\chi\neq\chi'}\right)$, enabling a matrix representation for the kernel $\boldsymbol{W}$, and likewise for the Liouvillian $\boldsymbol{L}$.
The generalized master equation, Eq.~(\ref{eq_master_general}), contains the full time dependence of the problem, which is affected by the time-dependent dot energy levels. In the limit when the lifetime of the double-dot states is much larger than the time scale given by the variation of the parameters, we can perform an adiabatic approximation along the lines of Ref.~\onlinecite{Splettstoesser06}. One first carries out a Taylor expansion, $\boldsymbol{P}\left(t'\right)=\sum_j \left(t'-t\right)^j/j!\frac{d^j}{dt^j}\boldsymbol{P}\left(t\right)$, around time $t$ of the matrix-elements of the reduced density matrix in the integrand of Eq.~(\ref{eq_master_general}), accounting for a finite memory of the kernel. In addition, the internal time dependence of the kernel via the time-dependent parameters is taken into account up to first order in $\Omega$. The adiabatic expansion of the kernel contains an instantaneous part $(i)$, where all parameters are fixed to their value at time $t$, $X(\tau)\rightarrow X(t)$. It further contains a first-order correction term $(a)$ which is found by considering $X(\tau)\rightarrow X(t)+(\tau-t)\frac{dX}{d\tau}|_{\tau=t}$ in the evaluation of the kernel and by taking into account systematically first-order terms in the time derivatives. This yields the two terms of the expansion
\begin{equation}
\boldsymbol{W}\left(t,t'\right)\rightarrow \boldsymbol{W}_t^{\left(i\right)}\left(t-t'\right)+\boldsymbol{W}_t^{\left(a\right)}\left(t-t'\right)\ .
\end{equation}
The resulting parametric time dependence is indicated by the subscript $t$. With this, a set of equations for the instantaneous and the adiabatic part, $\boldsymbol{P}_t^{\left(i\right)}$ and $\boldsymbol{P}_t^{\left(a\right)}$, of the reduced density matrix is constructed. The instantaneous contribution of the density matrix fulfills the stationary generalized master equation with system parameters frozen to values at time $t$. The adiabatic contribution accounts for the fact that the actual density matrix  slightly lags behind its instantaneous value.

We here restrict ourselves to the case of weak coupling. Therefore, on top of the \textit{adiabatic} expansion for small $\Omega$ we perform a \textit{perturbation} expansion in the \textit{tunnel coupling} between lead and dots. We take into account only terms up to first order in the tunnel coupling, which is valid if the broadening due to tunneling is smaller than the temperature broadening, $\Gamma< k_\mathrm{B} T$, where $k_\mathrm{B}$ is the Boltzmann constant. We perform an expansion of the generalized master equation order by order in $\Gamma$, as used in Ref.~\onlinecite{Konig96a,*Konig96b} and subsequent works. This systematic expansion ensures that all contributions to a given order are taken into account (it  may fail in certain situations when higher order terms in $\Gamma$ play a role,~\cite{Leijnse08} which is not the case in the regime considered here). 
We will also consider the case where $\Gamma$ is not the only small parameter but the intra-dot coupling $\Delta$ and the level difference $\epsilon$ have the same magnitude. Then coherences become important already in first order $\Gamma$, and the perturbation expansion has to be performed taking into account consistently all three small parameters, see Secs.~\ref{sec_weak_coupling} and~\ref{sec_spin_pump_weak}, as well as Appendix~\ref{app_pt_weak}. Gathering the terms of the lowest order in  $\Omega$ and the small parameters of the perturbative expansion, we end up with
\begin{subequations}
\begin{align}
0&=\left(\boldsymbol{W}_t^{\left(i,1\right)}-i\boldsymbol{L}_t^{\left(i,1\right)}\right)\boldsymbol{P}_t^{\left(i,0\right)}\label{eq_master_stat}\ , \\
\frac{d}{dt}\boldsymbol{P}_t^{\left(i,0\right)}&=\left(\boldsymbol{W}_t^{\left(i,1\right)}-i\boldsymbol{L}_t^{\left(i,1\right)}\right)\boldsymbol{P}_t^{\left(a,-1\right)}\ .
\label{eq_master_adiabatic}
\end{align}
\end{subequations}
Here we have introduced the zero-frequency Laplace transform $\boldsymbol{W}_t^{\left(i\right)}=\int_{-\infty}^{t}dt'\boldsymbol{W}_t^{\left(i\right)}\left(t-t'\right)$. 
The order in the perturbation expansion is indicated by numbers in the superscript of the respective quantities, i.e., first order in the transition elements $\boldsymbol{W}^{\left(i,1\right)}$, and the superscript of $\boldsymbol{L}^{\left(i,1\right)}$ indicates that the double-dot Liouvillian enters this equation in the same order of the Kernel and that its time dependence is instantaneous. The instantaneous as well as the adiabatic reduced density-matrix elements are $\boldsymbol{P}^{\left(i,0\right)}$ and $\boldsymbol{P}^{\left(a,-1\right)}$.
Note that the adiabatic correction to the reduced density matrix starts in minus first order in $\Gamma$; this is justified as these terms are proportional to $\Omega/\Gamma$, which is a small parameter in the adiabatic regime. Since we do not consider higher orders in the perturbation expansion, the corresponding superscript  for the above discussed quantities is unambiguous, and is suppressed in the following for simplicity.

Equations~(\ref{eq_master_stat}) and (\ref{eq_master_adiabatic}) have a very similar structure and therefore one can obtain Eq.~(\ref{eq_master_stat}) from Eq.~(\ref{eq_master_adiabatic}) by setting the left-hand side to zero and replacing the reduced density matrix by the instantaneous contribution. This will be used throughout the remainder of this paper.

A similar expansion is performed for the current through the system.
Generally, we can give the expression for the pumping current through the double quantum dot as
\begin{equation}\label{eq_current_general_time}
I_{\text{L}}\left(t\right)=e\,\boldsymbol{\text{e}}^{\text{T}}\int_{-\infty}^{t}dt' \boldsymbol{W}^{\text{L}}\left(t,t'\right)\boldsymbol{P}\left(t'\right)\ ,
\end{equation}
with the unity charge $e$. The trace of all contributing elements is performed by applying the vector $\boldsymbol{\text{e}}^{\text{T}}=(1,\ldots,1,0,\ldots,0)$, where the number of ones is equal to the dimensionality of the reduced Hilbert space. The kernel $\boldsymbol{W}^{\text{L}}\left(t,t'\right)$ with the zero-frequency Laplace transform $\boldsymbol{W}^\mathrm{L}$ takes into account processes which involve an exchange of charge between the left lead and the quantum dot subsystem, see Ref.~\onlinecite{Konig96a,*Konig96b}. 
In the case of zero bias, considered in this paper, the instantaneous contribution to the current, $I^{(i)}$, vanishes at all times and the only remaining term is the adiabatic contribution, $I^{(a)}$. In lowest (zeroth) order in the tunnel coupling we obtain
\begin{equation}\label{eq_current_general}
I_{\text{L}}(t):=I_{\text{L}}^{\left(a\right)}(t)=e\,\boldsymbol{\text{e}}^{\text{T}}\boldsymbol{W}_t^{\mathrm{L}\left(i\right)}\boldsymbol{P}_t^{\left(a\right)}\ ,
\end{equation}
where we drop the superscript $\left(a\right)$ for simplicity. Expressions (\ref{eq_current_general_time}) and (\ref{eq_current_general}) represent the tunneling current into the left lead. Note that due to the time dependence, additionally also displacement currents are present. In Ref.~\onlinecite{Bruder94} this has been studied in detail for a single dot. Importantly, the displacement current is a pure ac current and it averages out when integrating over one period of the pumping cycle $\mathcal{T}=2\pi/\Omega$. In the following we are mainly interested in the pumped charge per cycle, hence only the tunneling current plays a role and we discard the displacement current from now on. The number of pumped electrons is then given by
\begin{equation}\label{eq_N_general}
N=\frac{1}{e}\int_0^{\mathcal{T}}dt I_{\text{L}}\left(t\right)\ .
\end{equation}
We consider the system in a time-dependent steady state, i.e., no transient behavior; therefore the total charge of the double dot is conserved after one pumping cycle.
\footnote{In the following we will focus on the pumped charge for small pumping amplitudes on account of simplicity. Note however that this is no fundamental limitation, and the expression for the time-dependent current is valid also in the nonlinear regime. This regime of large pumping amplitudes becomes important if one is interested in quantized pumping. }

\section{Normal conductor reservoirs}\label{sec_N_dd_N}
In this section we discuss the pumped charge through the double-dot system in contact with two normal-conducting leads (N-DD-N).

\subsection{Weak interdot coupling}\label{sec_weak_coupling}

We first consider the situation where the two dots are weakly coupled to each other, $\Delta < k_\mathrm{B}T$, and the energy difference between a singly occupied left and a singly occupied right dot is small as well, $\epsilon <k_\mathrm{B}T$, such that both parameters are on the order of the coupling strength, $\Gamma\sim\Delta\sim\epsilon$. We account for expressions in the master equation up to first order in these parameters. In this case a rigorous expansion in the coupling parameters (up to lowest order in both lead-dot and dot-dot couplings) and the level difference $\epsilon$ is performed. This system has been considered in the \textit{static} case in Ref.~\onlinecite{Wunsch05}. For the remainder of this section we assume that not only $U'$ but also the charging energy $U$ is much larger than all other energy scales (such as temperature, the modulation frequency, and the level difference $\epsilon$). Therefore the double-dot system can only be singly occupied or empty. In this case the states $|\mathrm{L}\sigma\rangle$ and $|\mathrm{R}\sigma\rangle$, with the electron with spin $\sigma=\uparrow,\downarrow$ in the L or R dot, are almost-degenerate quasi eigenstates of the system and coherent superpositions of  $|\mathrm{L}\sigma\rangle$ and $|\mathrm{R}\sigma\rangle$ play an important role, see Appendix~\ref{app_pt_weak}. These coherent superpositions are captured in the off-diagonal elements of the reduced density matrix of the double-dot system. Even in the absence of bias, they do not vanish for an asymmetrically coupled double dot due to the time dependence of the system. The vector of the reduced density-matrix elements is $\mathbf{P}=(P_0,P_{\mathrm{L}\uparrow},P_{\mathrm{L}\downarrow},P_{\mathrm{R}\uparrow},P_{\mathrm{R}\downarrow},P_{\mathrm{L}\uparrow}^{\mathrm{R}\uparrow},P_{\mathrm{L}\downarrow}^{\mathrm{R}\downarrow},P_{\mathrm{R}\uparrow}^{\mathrm{L}\uparrow},P_{\mathrm{R}\downarrow}^{\mathrm{L}\downarrow})$, where we write the diagonal elements of the density matrix $P_{\chi}^{\chi}=:P_{\chi}$.
We end up with a master equation for the occupation probabilities for an empty dot $P_0$ and for a singly occupied dot $P_1=P_\mathrm{L}+P_\mathrm{R}$, where the total occupation with different spins is $P_\mathrm{L}=P_\mathrm{L\uparrow}+P_\mathrm{L\downarrow}$ and $P_\mathrm{R}=P_\mathrm{R\uparrow}+P_\mathrm{R\downarrow}$. We find
\begin{eqnarray}\label{eq_master1}
\frac{d}{ d t}P_{0} & = & -2\Gamma f^{+}\left(E\right)P_{0}+\vec{\mathrm{e}}_{\mathrm{z}}\lambda\Gamma f^{-}\left(E\right)\vec{\mathcal{S}} \nonumber\\
& &+\frac{1}{2}\Gamma f^{-}\left(E\right)P_{1},\
\end{eqnarray}
and $P_1$ is obtained via the probability conservation $P_0^{(i)}+P_1^{(i)}=1$ and $P_0^{(a)}+P_1^{(a)}=0$, for the instantaneous and the adiabatic parts of the reduced density-matrix elements. The tunnel coupling asymmetry is given by $\lambda=\left(\Gamma_{\text{L}}-\Gamma_{\text{R}}\right)/\Gamma$ taking values between $-1$ and $1$, where zero is the case of symmetric coupling to the leads.  The Fermi function is $f^+\left(E\right)=\frac{1}{e^{\beta E}+1}$ and $f^-\left(E\right)=1-f^+\left(E\right)$ with the inverse temperature $\beta=1/k_\mathrm{B} T$. 
The vector $\vec{\mathrm{e}}_\mathrm{z}$ projects out the $z$ component of the pseudospin vector which captures the off-diagonal elements and the difference in the occupation of the left and the right dots,
\begin{equation}
\vec{\mathcal{S}}=\frac{1}{2}\left(\begin{array}{c}
P_{\text{R}}^{\text{L}}+P_{\text{L}}^{\text{R}} \\
iP_{\text{R}}^{\text{L}}-iP_{\text{L}}^{\text{R}} \\
P_{\text{L}}-P_{\text{R}} \end{array}\right).
\end{equation}
The dynamics of the pseudospin is described by a Bloch-type equation 
\begin{align}
\begin{split}\label{eq_pseudospin}
\frac{ d}{ d t}\vec{\mathcal{S}}=&\,\vec{\mathrm{e}}_{\mathrm{z}}\lambda\Gamma\left(f^{+}\left(E\right)P_{0}-\frac{1}{4}f^{-}\left(E\right)P_{1}\right)\\ &-\frac{1}{2}\Gamma f^{-}\left(E\right)\vec{\mathcal{S}}+\vec{\mathcal{B}}\times\vec{\mathcal{S}} \  .
\end{split} 
\end{align}
The time evolution of the pseudospin has a contribution due to the accumulation of the pseudospin $z$ component, i.e., an unbalance of the left-right occupation, cf. first line of Eq. (\ref{eq_pseudospin}). Accumulation occurs only for asymmetric lead coupling. Relaxation of the pseudospin, given by the first contribution of the second line of  Eq. (\ref{eq_pseudospin}) takes place independently of the coupling asymmetry. Furthermore, we find a  precession of the pseudospin around an effective magnetic field, via which the internal dynamics of the double dot enter. It is given by
\begin{equation}\label{eq_pseudofield}
\vec{\mathcal{B}}=\left(\begin{array}{c}
-\Delta\\
0\\
\epsilon_{\text{ren}}\end{array}\right)\ .
\end{equation}
We note that instead of the bare level difference $\epsilon$, a renormalized one is entering the effective field, given by
\begin{equation}
\epsilon_{\text{ren}}=\epsilon+\lambda\frac{\Gamma}{2\pi}\phi\left(E\right)\ .
\end{equation}
This expression is explicitly time dependent via $\epsilon$ and $E$. The function $\phi$ entering the renormalized level difference is given by
\begin{equation}\label{eq_phi}
\phi\left(E\right)=\text{Re}\left[\psi\left(\frac{1}{2}+i\frac{\beta E}{2\pi}\right)\right]-\psi\left(\frac{1}{2}+\frac{\beta\epsilon_{\text{cutoff}}}{2\pi}\right)\ ,
\end{equation}
where $\psi$ denotes the digamma function. The renormalization arises from quantum charge fluctuations. It enters the $z$ component of the effective field, which acts as a Zeeman field for the pseudospin. This Zeeman field affects the pseudospin dynamics through a precession around the $z$ axis, thereby coupling to the $x$ and $y$ component of the pseudospin, which arise due to coherent superpositions of $|\mathrm{L}\sigma\rangle$ and $|\mathrm{R}\sigma\rangle$.
The level renormalization is a pure Coulomb interaction effect, i.e., it vanishes for $U=U'=0$. Furthermore, the renormalization is zero for symmetric coupling $\lambda=0$. 
Here, Coulomb interactions are large and $\epsilon_{\text{cutoff}}$ provides the cutoff energy. The full expression for the renormalization of $\epsilon$ for finite inter- and intra-dot Coulomb interaction is found in Ref.~\onlinecite{Wunsch05}. The effective field due to the interdot tunneling lies in the $x,y$-plane which is spanned by the pseudospin contribution due to the coherent superpositions of $|\mathrm{L}\sigma\rangle$ and $|\mathrm{R}\sigma\rangle$.

\begin{figure}
\centering
\includegraphics[scale=1]{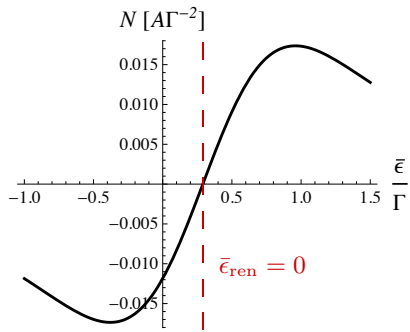}
\caption{(Color online.) Plot of the pumped charge $N$ with respect to the mean bare level difference $\bar{\epsilon}$ for $\bar{E}=1.5k_\mathrm{B}T$ and asymmetric tunnel coupling, $\lambda=1/2$. The node $N=0$ occurs when $\bar{\epsilon}_{\text{ren}}=0$. The other parameters are $\Delta=\Gamma$, $k_\mathrm{B}T=2\Gamma$, $\epsilon_{\text{cutoff}}=50k_\mathrm{B}T$. }
\label{fig_N_weak}
\end{figure}

The instantaneous solution of the master equation, Eqs. (\ref{eq_master1}) and (\ref{eq_pseudospin}), is given by Boltzmann distributions. The average instantaneous occupation number on the double dot $\langle n\rangle^{(i)}=0\cdot P_0+1\cdot P_1$ is given by  
\begin{equation}
\left\langle n \right\rangle^{(i)}=\frac{4e^{-\beta E}}{1+4e^{-\beta E}} \ .
\end{equation}
The factor four stems from the total fourfold degeneracy of the singly occupied state. For the same reason, we also have 
\begin{align}
\vec{\mathcal{S}}^{\left(i\right)}\left(t\right)&=0 \ ,
\end{align}
in lowest order in the tunnel couplings $\Gamma$ and $\Delta$ and the level difference $\epsilon$ (this holds as long as no bias voltage is present).
As opposed to the instantaneous solution, the adiabatic correction of the pseudospin $\vec{\mathcal{S}}^{(a,-1)}$ does in general not vanish.
This is due to an occupation difference in pseudospin space, introduced by the time dependence and asymmetric coupling to the left and right leads, $\lambda\neq0$.
Therefore, in the case of symmetric dot-lead coupling, the adiabatic correction for the pseudospin also vanishes. 
For the adiabatic current we find from Eq.~(\ref{eq_current_general}),
\begin{eqnarray}
I_{\text{L}}\left(t\right)&=&\frac{e}{2}\frac{ d \left\langle n \right\rangle^{(i)}}{ d t}\cdot \\ &&\left(1+\frac{\lambda\Delta^{2}}{\left(1-\lambda^2\right)\left(\frac{1}{4}\Gamma^{2}\left[f^{-}\left(E\right)\right]^{2}+\epsilon_{\text{ren}}^{2}\right)+\Delta^{2}}\right)\nonumber \ .
\end{eqnarray}
Note that the parameters $\epsilon$ and $E$ depend on time and that the pumping current is proportional to the time derivative of the occupation number, depending on $E$. Therefore, a necessary prerequisite for a nonzero adiabatic current is an explicitly time-dependent $E$.
Importantly, also the prefactor of $\frac{ d}{ d t}\left\langle n\right\rangle^{\left(i\right)}$ is in general time dependent. This is the necessary condition for a non-vanishing average pumped charge.
\footnote{When considering the total current pumped out of the dot through the left and the right lead, we recover the total displacement current $e\frac{ d}{ d t}\protect\langle n\protect\rangle^{(i)}$, as required.}\\
If $\lambda=0$ we find that the time-dependent current through each contact $\alpha=\mathrm{L,R}$ is given by $I_\alpha(t)=\frac{e}{2}\frac{ d}{ d t}\left\langle n\right\rangle^{\left(i\right)}$. Therefore  the currents injected into the left and the right leads are equal at any time and the pumped charge per cycle vanishes; it is therefore directly sensitive to the coupling asymmetry.\\
The pumped charge is obtained from Eq.~(\ref{eq_N_general}). 
We are interested in the  regime of bilinear response for the modulation, which is valid if the pumped charge per infinitesimal area in parameter space varies little within the area enclosed by the cycle. Then the pumped charge is proportional to the cycle area, $A=\int_{0}^{\mathcal{T}}dt\delta\epsilon\delta\dot{E}$. We find  the number of pumped charges $N$ per area in parameter space as a function of the average quantities $\bar{\epsilon}$ and $\bar{E}$,
\begin{equation}\label{eq_charge_weak}
\frac{N}{A}=\frac{-\lambda\left(1-\lambda^2\right)\bar{\epsilon}_{\text{ren}}\Delta^{2}}{\left(\left(1-\lambda^2\right)\left(\left[\frac{\Gamma}{2}f^{-}\left(\bar{E}\right)\right]^{2}+\bar{\epsilon}_{\text{ren}}^{2}\right)+\Delta^{2}\right)^{2}}\frac{ d  \bar{\left\langle n \right\rangle}^{(i)}}{ d \bar{E}}\ .
\end{equation}
The pumped charge has a peak when $\bar{E}$ is close to resonance and fulfills the relations
\begin{subequations}
\begin{align}
N\left(\bar{\epsilon}_{\text{ren}}\right)=&-N\left(-\bar{\epsilon}_{\text{ren}}\right)\ , \\
N\left(\lambda\right)=&-N\left(-\lambda\right)\ .
\end{align}
\end{subequations}
Inverting both parameters consequently maps the function $N$ onto itself again. 
As discussed above, the pumped charge vanishes if $\lambda=0$, as well as at $\lambda=\pm1$ when one of the leads is completely decoupled. 
Since the pumped charge also vanishes if $\bar{\epsilon}_{\text{ren}}=0$ (see Fig.~\ref{fig_N_weak}), the level renormalization can be directly read out by means of pumping through the double dot, when scanning through the time-averaged left-right level difference $\bar{\epsilon}$. This important property occurs due to the following reason:
the prefactor of $\frac{ d}{ d t}\left\langle n\right\rangle^{\left(i\right)}$ is an even function of the renormalized difference of the left and right level position $\epsilon_\mathrm{ren}(t)$; therefore in the limit of small pumping amplitudes (bilinear response, i.e., only the time dependence of $\epsilon$ matters in the prefactor) a sign change in the average  $\bar{\epsilon}_{\text{ren}}$ has the same effect as a shift of the modulation by $\pi$. Doing the average over one pumping period, the transport direction is therefore reversed.\\
The level renormalization due to quantum charge fluctuations, which can be measured in the pumped charge, is distinguishable from possible level renormalization effects due to an energy-dependent density of states  by its temperature dependence, which is logarithmic for large $k_\mathrm{B}T$.	

\subsection{Strong interdot coupling}\label{sec_strong_coupling}
A finite interdot coupling amplitude $\Delta$ leads to a hybridization of the energy levels of the right and the left dots. In the limit of strong coupling between the two dots, $\Delta>\Gamma$, we write the single-particle part of the double-dot Hamiltonian 
in terms of bonding $|\mathrm{b}\rangle$ and antibonding $|\mathrm{a}\rangle$ states, $\sum_{\eta\sigma}\epsilon_{\eta}\left(t\right)d_{\eta\sigma}^{\dagger}d_{\eta\sigma}$, where $\eta=\mathrm{b,a}$ and $\sigma=\uparrow,\downarrow$.
The energy of the bonding and antibonding state is given by
\begin{equation}
\epsilon_{\text{b}/\text{a}}=E\mp\frac{1}{2}\sqrt{\Delta^{2}+\epsilon^{2}}.
\end{equation}
The corresponding operators creating (annihilating) an electron in the bonding and the antibonding states are related to the ones creating an electron in the left or the right dot by the equality
\begin{equation}
d_{\text{L}/\text{R}}^{\dagger}=\frac{1}{\sqrt{2}}\sqrt{1\mp\frac{\epsilon}{\sqrt{\Delta^{2}+\epsilon^{2}}}}d_{\text{b}}^{\dagger}\mp\frac{1}{\sqrt{2}}\sqrt{1\pm\frac{\epsilon}{\sqrt{\Delta^{2}+\epsilon^{2}}}}d_{\text{a}}^{\dagger}\ ,
\end{equation}
where we omitted the spin indices for simplicity. The bonding $|\mathrm{b}\rangle$ and antibonding $|\mathrm{a}\rangle$ states have an energy difference much larger than the level broadening. This means that the interdot dynamics is much faster than the dot-lead hopping.
Therefore, coherent superpositions of 
bonding and antibonding states are suppressed in lowest order in the tunneling and 
no dynamics of the off-diagonal elements of the reduced density matrix have to be considered (in contrast to the previous case discussed in Sec.~\ref{sec_weak_coupling}). 
We now take $U$ to be finite and we therefore also consider the  occupation of the left and right dots  with one electron each. We still assume large on-site Coulomb interaction, inhibiting double occupation of each single dot. As a consequence, the doubly occupied (fourfold degenerate) states are $\left|\mathrm{L}\sigma \mathrm{R}\sigma'\right\rangle$.  We can now write the double-dot Hamiltonian $\sum_\chi E_\chi|\chi\rangle\langle\chi|$ in the basis of the eigenstates $|\chi\rangle$. Here the eigenenergy $E_0$ of the empty dot $|0\rangle$ equals zero, the eigenenergy $E_\mathrm{b}$ of the spin-degenerate bonding states $|\mathrm{b}\sigma\rangle=d^\dagger_{\mathrm{b}\sigma}|0\rangle$ is $\epsilon_\mathrm{b}$ and analogously the eigenenergy $E_\mathrm{a}$ of the spin-degenerate antibonding states $|\mathrm{a}\sigma\rangle=d^\dagger_{\mathrm{a}\sigma}|0\rangle$ is $\epsilon_\mathrm{a}$. Finally the four doubly occupied states $\left|\mathrm{L}\sigma \mathrm{R}\sigma'\right\rangle$, with $\sigma=\uparrow,\downarrow$ and $\sigma'=\uparrow,\downarrow$, have the eigenenergies $E_{\sigma\sigma'}=2E+U$.
The tunnel coupling between the double dot and the leads $\alpha=\mathrm{L,R}$ is captured via effective rates for tunneling through the hybrid states $|\mathrm{b}\rangle$ and $|\mathrm{a}\rangle$. These rates are explicitly time dependent and are given by
\begin{equation}
\label{eq_eff_Gamma}
\Gamma_{\alpha\eta}=\frac{1}{2}\left(1-\alpha\eta\frac{\epsilon}{\sqrt{\Delta^{2}+\epsilon^{2}}}\right)\Gamma_{\alpha}\ .
\end{equation}
Tunneling through the hybrid single-particle states is denoted by the subscript $\eta=\mathrm{b,a}$. The difference of the coupling strengths for the two transport channels is proportional to  a factor of $\epsilon/\sqrt{\Delta^{2}+\epsilon^{2}}$. (This factor is related to the relative position of the bonding and antibonding states with respect to the localized states $\mathrm{L}$ and $\mathrm{R}$.) In order to bring out this important property we use the following notation: if $\eta$ is used as a variable rather than a coefficient it takes the value $+1$ ($-1$) for b (a); equally if $\alpha=\mathrm{L,R}$ is used as a variable it takes the values $+1$ for L and $-1$ for R.  The sum of these rates is $\Gamma_{\eta}=\sum_{\alpha}\Gamma_{\alpha\eta}$. 
With this, the master equation for the occupation probabilities $P_0$, $P_\eta=P_{\eta\uparrow}+P_{\eta\downarrow}$, and $P_\mathrm{d}=\sum_{\sigma,\sigma'=\uparrow,\downarrow}P_{\mathrm{L}\sigma\mathrm{R}\sigma'}$ reads
\begin{subequations}
\begin{eqnarray}
\frac{d}{dt}P_{0}&=&\sum_{\eta=\text{b},\text{a}}\left(-2\Gamma_{\eta}f^{+}\left(\epsilon_{\eta}\right)P_{0}+\Gamma_{\eta}f^{-}\left(\epsilon_{\eta}\right)P_{\eta}\right)\ , \\
\frac{d}{dt}P_{\eta}&=&\, 2\Gamma_{\eta}f^{+}\left(\epsilon_{\eta}\right)P_{0}+\left(-\Gamma_{\eta}f^{-}\left(\epsilon_{\eta}\right)\right. \\ &&\left.\phantom{\frac{1}{2}}-2\Gamma_{\bar{\eta}}f^{+}\left(\epsilon_{\bar{\eta}}+U\right)
\right)P_{\eta}+\Gamma_{\bar{\eta}}f^{-}\left(\epsilon_{\bar{\eta}}+U\right)P_\mathrm{d}\nonumber\ ,
\end{eqnarray}
\end{subequations}
and $P_\mathrm{d}$ is obtained via the probability conservation $P_0^{(i)}+P_\mathrm{b}^{(i)}+P_\mathrm{a}^{(i)}+P_\mathrm{d}^{(i)}=1$ and $P_0^{(a)}+P_\mathrm{b}^{(a)}+P_\mathrm{a}^{(a)}+P_\mathrm{d}^{(a)}=0$. 
The instantaneous probabilities are again given by the Boltzmann distribution. \\
As described before, in Sec. \ref{sec_diagram}, the instantaneous occupation probabilities and their first-order corrections in the adiabatic expansion are found as solutions of this master equation.
The adiabatic current is then calculated straightforwardly as
\begin{equation}
I_{\text{L}}\left(t\right)=e\sum_{\eta=\mathrm{b,a}}\frac{\Gamma_{\text{L}\eta}}{\Gamma_{\eta}}\left(\frac{ d}{ d t}P_{\eta}^{\left(i\right)}+\frac{ d}{ d t}P_\mathrm{d}^{\left(i\right)}\right).
\end{equation}
The current expression is divided into two different contributions, $I_{\text{L}}\left(t\right)=I_{\text{L,b}}\left(t\right)+I_{\text{L,a}}\left(t\right)$. It is apparent that the dot state transitions $|0\rangle\leftrightarrow|\mathrm{b}\rangle$ and $|\mathrm{a}\rangle\leftrightarrow|\mathrm{d}\rangle$ occur with the rate $\Gamma_\mathrm{b}$. And likewise, the transitions $|0\rangle\leftrightarrow|\mathrm{a}\rangle$ and $|\mathrm{b}\rangle\leftrightarrow|\mathrm{d}\rangle$ are due to tunneling with the rate $\Gamma_\mathrm{a}$.

\begin{figure}
\centering
\includegraphics[scale=1]{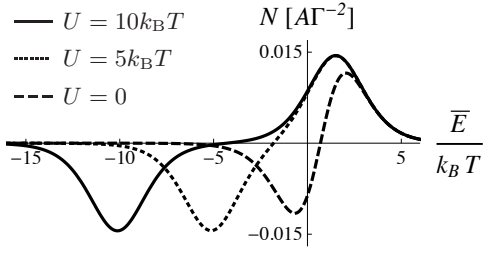}
\caption{Plot of the pumped charge $N$ as a function of $\bar{E}$ for different values for the Coulomb interaction $U={0,5k_\mathrm{B}T,10k_\mathrm{B}T}$ (dashed, dotted, solid), in the symmetric case $\bar{\epsilon}=0$, $\lambda=0$, and the interdot coupling is $\Delta=k_\mathrm{B}T$ and $k_BT=2\Gamma$. }
\label{fig_N_ba_U}
\end{figure}

In bilinear response for small pumping amplitudes, the pumped charge through the double dot becomes
\begin{equation}\label{eq_N_strong_Delta}
\frac{N}{A}=\sum_{\eta=\mathrm{b,a}}\frac{ d}{ d \bar{\epsilon}}\left(\frac{\bar{\Gamma}_{\text{L}\eta}}{\bar{\Gamma}_{\eta}}\right)\frac{ d}{ d \bar{E}}\left(\bar{P}_{\eta}^{\left(i\right)}+\bar{P}_\mathrm{d}^{\left(i\right)}\right)\ .
\end{equation}
The derivative with respect to $\bar{E}$ of the sum of the 	probabilities is always negative for both the bonding and the antibonding contributions. In contrast, the prefactor, namely, the derivative with respect to $\bar{\epsilon}$ of the relative effective coupling, has opposite signs for the different hybrid states, since the time-dependent part of the effective coupling comes always with opposite signs, respectively, see Eq.~(\ref{eq_eff_Gamma}). 
  
As a direct result, one finds that the terms $N_\mathrm{b}$ and $N_\mathrm{a}$, derived from the current separated into $I_{\text{L,b}}\left(t\right)$ and $I_{\text{L,a}}\left(t\right)$, always contribute to the pumped charge per area with \textit{opposite} signs. 
This result is a central point of our paper; this double-dot feature cannot be provided by a single-dot system and is of fundamental importance for many of the effects discussed in the following.

In Fig.~\ref{fig_N_ba_U} we plot the pumped charge as a function of the mean level $\bar{E}$ in the case of spatial symmetry $\bar{\epsilon}=0$, $\lambda=0$ and for different values of $U$. Importantly, the finite Coulomb interaction shifts the resonance positions and furthermore enhances the pumped charge.
The two resonant peaks have opposite signs; they appear when the addition energy to go from an empty dot to the ground state of the singly occupied dot is at resonance, $\bar{\epsilon}_\mathrm{b}\approx0$ and when the addition energy to go from the singly occupied ground state to a doubly occupied dot is at resonance, $\bar{\epsilon}_\mathrm{a}+U\approx 0$, except for the common temperature-dependent shift due to different charging and decharging rates.
Since the main contribution at  $\bar{\epsilon}_\mathrm{b}\approx0$ comes from tunneling involving the hybrid state $|\mathrm{b}\rangle$ and the main contribution at  $\bar{\epsilon}_\mathrm{a}+U\approx0$ comes from tunneling involving the hybrid state $|\mathrm{a}\rangle$, the pumped charge has opposite signs at the two resonances. The pumped charge from the two contributions does not cancel out for $U=0$ because of the finite level spacing. 
This result agrees with Refs.~\onlinecite{Buitelaar08a,Kashcheyevs08}, where charge pumping through a double dot with noninteracting spinless electrons was considered.

If the double-dot parameters are such that the level splitting $|\bar{\epsilon}_{\text{b}}-\bar{\epsilon}_{\text{a}}|$ is much larger than $k_\mathrm{B}T$, the probability of having state $\left|\text{a}\right\rangle$ occupied vanishes. Thus, the transport close to the single-electron resonance  occurs solely via charging and decharging the single-electron ground state $\left|\text{b}\right\rangle$, via the processes $\left|0\right\rangle\leftrightarrow\left|\text{b}\right\rangle$, contributing with rate $\Gamma_\mathrm{b}$. 
Equally, at the resonance for the transition between singly and doubly occupied double dots, only the transitions $\left|\mathrm{b}\right\rangle\leftrightarrow\left|\text{d}\right\rangle$ contribute with rate $\Gamma_\mathrm{a}$.

\begin{figure}
\centering
\includegraphics[scale=1]{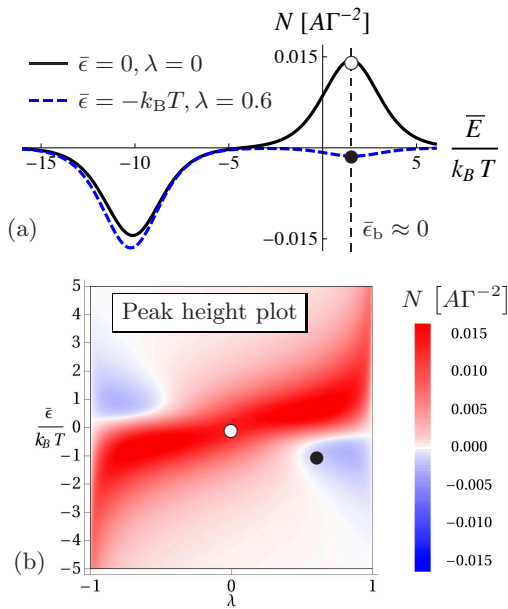}
\caption{(Color online.) (a) Plot of $N$ as a function of $\bar{E}$ for the asymmetric case (blue dashed) and the symmetric case taken from Fig.~\ref{fig_N_ba_U} as reference (black solid). (b) Map of the peak height of the pumped charge $N$ at the single-electron resonance (dashed vertical line in (a)) in dependence of $\bar{\epsilon}$ and $\lambda=\left(\Gamma_{\text{L}}-\Gamma_{\text{R}}\right)/\Gamma$. The empty and filled circle indicate the symmetry configurations taken in (a). For all plots the interdot coupling is $\Delta=k_\mathrm{B}T$ and the Coulomb interaction is $U=10k_\mathrm{B}T$. The thermal energy is $k_BT=2\Gamma$.}
\label{fig_N_ba}
\end{figure}

In the regime when the excited state is still thermally accessible, $\left|\bar{\epsilon}_{\text{b}}-\bar{\epsilon}_{\text{a}}\right|\lesssim k_\mathrm{B}T$, also the transitions with the excited state, $\left|0\right\rangle\leftrightarrow\left|\text{a}\right\rangle$ and $\left|\mathrm{a}\right\rangle\leftrightarrow\left|\text{d}\right\rangle$, start to contribute. By introducing a spatial asymmetry to the double dot, choosing $\bar{\epsilon}\neq0$ or $\lambda\neq0$, the two contributions  (for transport involving the ground state and transport involving the thermally excited state) are changed by different amounts. 
In spite of the fact that  the larger time-dependent current flow stems from transport through the ground state, for the pumped charge, i.e., the time-averaged quantity, the contribution through the ground state can become the minor one. This means that by choosing appropriate double-dot parameters charging and decharging of the ground state leads to smaller directed transport than charging and decharging of the excited state, a unique feature of time-averaged transport due to time-dependent fields.
Thereby, we can achieve a sign change in the pumped charge at one of the resonances, see Fig.~\ref{fig_N_ba}(a), (analogously for the other resonance at $\bar{\epsilon}_\mathrm{a}+U\approx0$ a sign change appears for $\lambda\rightarrow-\lambda$ or $\epsilon\rightarrow-\epsilon$). 

We plot the peak height of the pumped charge close to the resonance $\bar{\epsilon}_{\text{b}}\approx0$ for the full parameter spectrum of $\lambda$ and $\bar{\epsilon}$ in Fig.~\ref{fig_N_ba}(b). There, we see that for most parameter configurations, the bonding channel is dominant hence the peak is positive. The blue regions (brighter areas in the lower  right and the upper left) show the parameter regimes where a sign change appears due to the dominant transport through the thermally accessible antibonding state. These sign changes in the peaks with changing asymmetry will however vanish, once the level splitting becomes significantly larger than $k_\mathrm{B}T$ and the excited state is no longer thermally accessible.
We also find that the pumped charge is point symmetric with respect to the parameter $\lambda$ and the bare, time-averaged $\bar{\epsilon}$, $N\left(\lambda,\bar{\epsilon}\right)=N\left(-\lambda,-\bar{\epsilon}\right)$.

\section{Spin pumping through N-DD-F}\label{sec_spin_pump}

We now replace one of the contacts by a ferromagnetic lead (we choose the right one), thereby breaking spin-rotation invariance which enables spin in addition to charge pumping. Spin and charge pumping through a \textit{single} interacting quantum dot in the presence of a ferromagnetic lead has been studied before;~\cite{Splettstoesser08a} spin pumping through  a \textit{double} dot is particularly promising due to the following reason: the possible sign reversal of the charge transport which in the weak-coupling regime is taking place at the renormalized level difference $\bar{\epsilon}_\mathrm{ren}$ being zero, and which in the strong-coupling case is due to thermal accessibility of the excited level, is expected to affect the transport for different spin differently. \\
We now discuss the representative situation at the resonance between empty and singly occupied double dots and therefore restrict the calculation to infinite Coulomb interactions $U$ and $U'$ for the remainder of this paper. In a spin-polarized contact $\alpha$ the density of states is spin dependent, $\rho_{\alpha\uparrow}\neq\rho_{\alpha\downarrow}$. The spin polarization strength of lead $\alpha$ is defined as
\begin{equation}\label{eq_p_alpha}
p_{\alpha}=\frac{\rho_{\alpha\uparrow}-\rho_{\alpha\downarrow}}{\rho_{\alpha\uparrow}+\rho_{\alpha\downarrow}}.
\end{equation}
Due to the spin-dependent density of states also the tunneling rates to  the right lead become spin dependent
\begin{align}
\Gamma_{\alpha\sigma}=&\left(1+\sigma p_\alpha\right)\Gamma_{\alpha}\ ,
\end{align}
where $\sigma=+1$ for $\Gamma_{\alpha\uparrow}$ and  $\sigma=-1$ for $\Gamma_{\alpha\downarrow}$ and $\alpha=\mathrm{L,R}$. The total tunneling rate for lead $\alpha$ is $\Gamma_\alpha=\frac{1}{2}\left(\Gamma_{\alpha\uparrow}+\Gamma_{\alpha\downarrow}\right)$.
In this section only $p_\mathrm{R}$ is different from zero hence $\Gamma_{\mathrm{L}\sigma}=\Gamma_{\mathrm{L}}$. We choose the polarization axis of the spin in all parts of the N-DD-F system along the axis of the majority spin of the ferromagnetic contact and we can study the dynamics of electrons with spin up and spin down separately. 
Due to the spin-dependent tunneling rates, the spin-up and the spin-down channels have different pumping dynamics, leading to a generally nonzero net spin transport. Analogously to Sec.~\ref{sec_N_dd_N}, we discuss in the following the two limits for weak and strong interdot coupling. We find that the equations for the spin-resolved pumped charge are formally equivalent to the results of the unpolarized case, see Sec.~\ref{sec_N_dd_N}, where one  replaces $\Gamma_{\alpha}\rightarrow\Gamma_{\alpha\sigma}$. We will show that the two previously discussed different regimes both lead to a pure pumped spin current, relying on different effects.

\subsection{Weak interdot coupling}\label{sec_spin_pump_weak}

We first discuss the case of weak interdot coupling, where $\Delta\sim\Gamma\sim\epsilon$, analogous to Sec.~\ref{sec_weak_coupling}.
In order to calculate the total pumped charge $N=N_\uparrow+N_\downarrow$  and the total pumped spin $N^S=N_\uparrow-N_\downarrow$, we evaluate the number of pumped electrons with spin $\sigma$. We obtain an expression similar to the unpolarized  case from Eq.~(\ref{eq_charge_weak}),
\begin{eqnarray}
&&\frac{N_\sigma}{A}=-\frac{1}{2}\frac{ d\bar{\left\langle n \right\rangle}^{(i)}}{ d \bar{E}}\\
& &\times \frac{\lambda_\sigma\left(1-{\lambda_\sigma}^2\right)\bar{\epsilon}_{\text{ren},\sigma}\Delta^{2}}{\left(\left(1-{\lambda_\sigma}^2\right)\left(\left[\frac{\Gamma_{\mathrm{L}\sigma}+\Gamma_{\mathrm{R}\sigma}}{2} f^{-}\left(\bar{E}\right)\right]^{2}+{\bar{\epsilon}_{\text{ren},\sigma}}^{2}\right)+\Delta^{2}\right)^{2}}\ ,
 \nonumber
\end{eqnarray}
with the spin-dependent quantities
\begin{subequations}
\begin{eqnarray}
\lambda_\sigma & = &   \frac{\Gamma_{\text{L}\sigma}-\Gamma_{\text{R}\sigma}}{\Gamma_{\text{L}\sigma}+\Gamma_{\text{R}\sigma}}\ , \\ 
\epsilon_{\text{ren},\sigma} & = & \epsilon+\frac{\left(\Gamma_{\mathrm{L}\sigma}+\Gamma_{\mathrm{R}\sigma}\right)\lambda_{\sigma}}{2\pi}\phi\left(E\right)\ .
\end{eqnarray}
\end{subequations}
Importantly, the coupling asymmetry is spin dependent here, more specifically, for the spin-dependent coupling asymmetries we always find $\lambda_\uparrow\leq\lambda\leq\lambda_\downarrow$. This also makes the renormalized energy-level distance $\epsilon_{\text{ren},\sigma}$ spin dependent. As we have elaborated in Sec.~\ref{sec_weak_coupling}, there are nodes in the pumped charge whenever the left and right tunnel rates are equal, and when the time-averaged renormalized level difference $\bar{\epsilon}_\mathrm{ren}$ is zero. In analogy, we find nodes for the spin-resolved pumped charge $N_\sigma$:  zero net transfer of particles with spin $\sigma$ occurs if  $\lambda_{\sigma}=0$ and if  $\bar{\epsilon}_{\text{ren},\sigma}=0$. 

\begin{figure}
\centering
\includegraphics[scale=1]{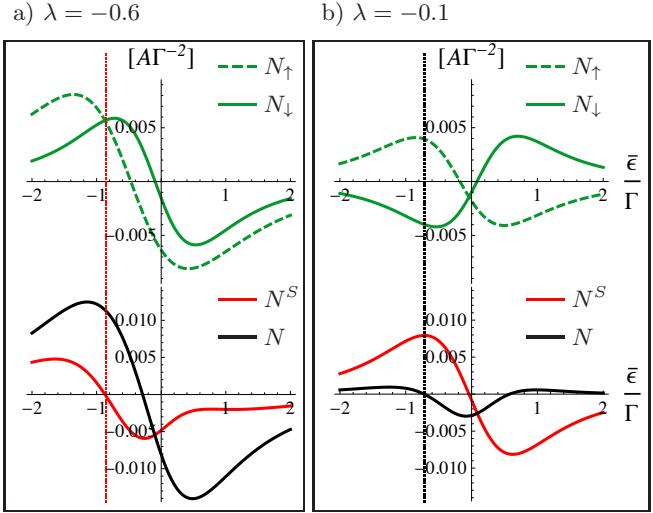}
\caption{(Color online.) Plot of the pumped charge through the up spin state $N_\uparrow$ and down spin state $N_\downarrow$ (upper plot), and the total pumped charge $N=N_\uparrow+N_\downarrow$ as well as the total pumped spin $N^S=N_\uparrow-N_\downarrow$ (lower plot), for different coupling symmetries (a) $\lambda=-0.6$ and (b) $\lambda=-0.1$. The polarization strength of the right reservoir is $p_{\text{R}}=0.5$.  The other parameters are $\Delta=\Gamma$, $\bar{E}=1.5k_\mathrm{B}T$ and $k_\mathrm{B}T=2\Gamma$.}
\label{fig_N_S_weak}
\end{figure}
The spin-resolved pumped charges $N_\uparrow$ and $N_\downarrow$ are plotted in the upper panels of Fig.~\ref{fig_N_S_weak} for different values of the coupling asymmetry. Clearly, the number of pumped charges with spin up and down differs strongly from each other leading to a finite spin transport. If the coupling asymmetry is such that $\Gamma_{\mathrm{R}\downarrow}<\Gamma_{\mathrm{L}}<\Gamma_{\mathrm{R}\uparrow}$, then spin-up and spin-down electrons are even mostly pumped in opposite directions as is shown in the upper panel of Fig.~\ref{fig_N_S_weak}~(b).

The lower panels of Fig.~\ref{fig_N_S_weak} show the total pumped charge and the total pumped spin. We see that the pumped charge and spin have different nodes: whenever $N_\uparrow=N_\downarrow$ the pumped spin vanishes, as indicated by the vertical dotted line in Fig.~\ref{fig_N_S_weak} (a), while the pumped charge can still be finite. On the other hand, when $N_\uparrow=-N_\downarrow$ the pumped charge vanishes, as indicated by the vertical dotted line in Fig.~\ref{fig_N_S_weak} (b), while the pumped spin can still be finite. This results in both the possibility to pump charge in the absence of net spin transport, and quite more intriguing, the possibility to pump spin without net charge transport by electrical control only.
In the regime studied here, pure spin pumping in the absence of charge pumping is due to spin-dependent quantum charge fluctuations, induced by Coulomb interactions, resulting in differing spin-dependent level renormalization effects.

\subsection{Strong interdot coupling}

\begin{figure}[t]
\centering
\includegraphics[scale=1]{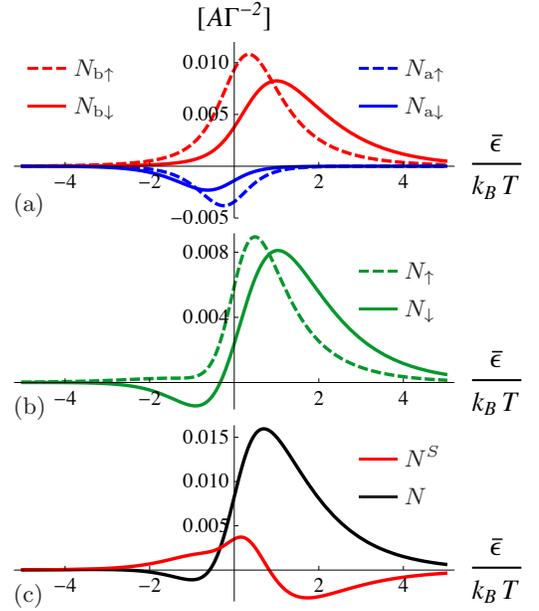}
\caption{(Color online.) (a) Plot of the pumped charge resolved for all four possible channels $N_{\eta\sigma}$, $\eta=\mathrm{b,a}$ and $\sigma=\uparrow,\downarrow$. (b) Spin-resolved pumped charge, $N_\uparrow$ and  $N_\downarrow$. (c) The total pumped charge $N=N_\uparrow+N_\downarrow$ as well as the total pumped spin $N^S=N_\uparrow-N_\downarrow$. For all plots the coupling symmetry is $\lambda=0.6$, the polarization strength of the right reservoir is $p_{\text{R}}=0.5$, and the other parameters are $\Delta=k_\mathrm{B}T$, $\bar{E}=1.5k_\mathrm{B}T$ and $k_\mathrm{B}T=2\Gamma$.}
\label{fig_N_S_strong}
\end{figure}

We now turn to the strong interdot coupling regime, $\Delta>\Gamma$, which is studied in Sec.~\ref{sec_strong_coupling} for unpolarized leads. Similar to Eq.~(\ref{eq_N_strong_Delta}), we obtain for the spin-resolved pumped charge through the two states $\eta=\mathrm{b,a}$,
\begin{equation}
\frac{N_{\eta\sigma}}{A}=\frac{1}{2}\frac{ d}{ d\bar{\epsilon}}\left(\frac{\bar{\Gamma}_{\text{L}\eta}}{\bar{\Gamma}_{\eta}+\sigma p_{\text{R}}\bar{\Gamma}_{\text{R}\eta}}\right)\frac{ d \bar{P}_{\eta}^{\left(i\right)}}{ d \bar{E}}\ .
\end{equation}
These quantities depend explicitly on the spin and the state $\eta$ through which the charge is pumped and therefore all four quantities are in general expected to be different, leading to possible pure dc charge as well as pure dc spin transport, see Fig.~\ref{fig_N_S_strong}(c). 

To explain the origin of different $N_\uparrow$ and $N_\downarrow$ we consider Figs.~\ref{fig_N_S_strong}(a) and \ref{fig_N_S_strong}(b). In Fig.~\ref{fig_N_S_strong}(a) we resolve the pumped charge for each channel separately, namely, for a spin-up electron through both hybrid states, $N_{\text{b}\uparrow}$ and $N_{\text{a}\uparrow}$, and likewise for a spin-down electron, $N_{\text{b}\downarrow}$ and $N_{\text{a}\downarrow}$.  For the plots in Fig.~\ref{fig_N_S_strong} we take $\bar{\epsilon}_\mathrm{b}\approx0$ in the case of strong coupling asymmetry $\lambda=0.6$ where we consider $\Delta$ and $\bar{\epsilon}$ such that the excited state (antibonding)  is  thermally accessible. 
Figure~\ref{fig_N_S_strong}(a) shows contributions due to transport through the bonding state (red), which have always an opposite sign to the charge pumped through the antibonding state (blue). Since the spin polarization enters Eq.~(\ref{eq_eff_Gamma}) only as a modification of the prefactor $\Gamma_\mathrm{R}$, the important sign dependence of the time-dependent part of the effective tunnel coupling for different hybrid channels  is not altered. Consequently, as discussed in Sec.~\ref{sec_strong_coupling}, also the spin-resolved pumped charge has opposite signs for  different hybrid channels. Whether the charge pumped through the excited, antibonding level can dominate over the contribution through the bonding state, depends strongly on the coupling asymmetry to the leads. In the example shown here, for the spin-down channel the increased effective coupling asymmetry allows for dominant transport through the antibonding hybrid state. This leads to a sign change in the pumped charge with spin down $N_\downarrow$ as a function of $\bar{\epsilon}$, in the limit of a thermally accessible excited state, see Fig.~\ref{fig_N_S_strong}(b).
In contrast, there is no configuration in which the transport of electrons with spin up takes place prevalently through the excited state since $\lambda_\uparrow<\lambda$ is too small to invert the transport direction.
The important result is that $N_\uparrow$ and $N_\downarrow$ can have opposite signs for certain values of $\bar{\epsilon}$. Therefore we find again the possibility of \textit{pure} spin pumping while the pumped charge vanishes and vice versa.

\section{Noncollinear ferromagnetic reservoirs}\label{sec_ferro}

\begin{figure}
\centering
\includegraphics[scale=1]{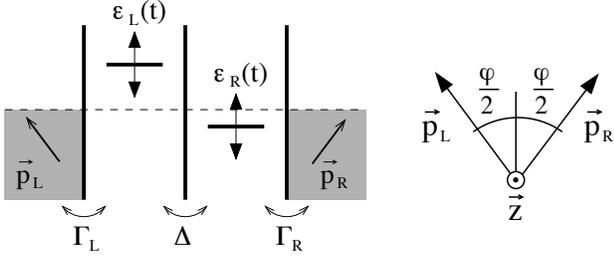}
\caption{A schematic picture of the Hamiltonian with ferromagnetic leads. The magnetization of the leads is given by $\vec{p}_{\text{L}}$ and $\vec{p}_{\text{R}}$.}
\label{fig_ferro}
\end{figure}

In this section we discuss the pumping characteristics through the double-dot device coupled to two differently polarized ferromagnetic leads, F-DD-F, as sketched in Fig.~\ref{fig_ferro}. We concentrate on the strong interdot coupling limit $\Delta>\Gamma$ and  start from the Hamiltonian discussed in Sec.~\ref{sec_hamiltonian}; it is now important to keep track of the different tunnel coupling for minority and majority spin of the different reservoirs.

The polarization strength $p_\alpha$, Eq.~(\ref{eq_p_alpha}), and the polarization direction defining the polarization axes  $\vec{n}_\alpha$ are different for the two leads $\alpha=\mathrm{L,R}$. We choose the spin quantization axis of each lead along the respective magnetization axis, $\vec{n}_{\text{L}}$ and $\vec{n}_{\text{R}}$, where the two vectors enclose an angle $\varphi$. 
Following Ref.~\onlinecite{Braun04}, we now take a coordinate system such that its basis vectors $\hat{e}_x$, $\hat{e}_y$, and $\hat{e}_z$ align with $\vec{n}_{\text{L}}+\vec{n}_{\text{R}}$, $\vec{n}_{\text{L}}-\vec{n}_{\text{R}}$, and $\vec{n}_{\text{L}}\times\vec{n}_{\text{R}}$, respectively. For the dots' spin quantization axis we choose the $z$ axis of this coordinate system, which is perpendicular to the plane spanned by the leads' polarization vectors, see Fig.~\ref{fig_ferro}. In this basis, the left and right lead magnetization vectors, $\vec{p}_\mathrm{L}=p_\mathrm{L}\vec{n}_\mathrm{L}$ and $\vec{p}_\mathrm{R}=p_\mathrm{R}\vec{n}_\mathrm{R}$, are given as
\begin{equation}
\vec{p}_{\text{L}}=p_{\text{L}}\left(\begin{array}{c}
\cos\left(\varphi/2\right)\\
\sin\left(\varphi/2\right)\\
0\end{array}\right), \ \vec{p}_{\text{R}}=p_{\text{R}}\left(\begin{array}{c}
\cos\left(\varphi/2\right)\\
-\sin\left(\varphi/2\right)\\
0\end{array}\right)\ .
\end{equation}
It is useful to express the tunneling Hamiltonian, Eq.~(\ref{eq_tunnel_ham}), in terms of creation (annihilation) operators for an electron with a spin along the spin quantization axis of the respective system part. We therefore write
\begin{equation}
\begin{split}
H_{\text{tunnel}}=&\sum_{\alpha=\text{L},\text{R}}\frac{V_{\alpha}}{\sqrt{2}}\sum_{k}\left\{c_{\alpha k+}^{\dagger}\left(e^{i\alpha\varphi/4}d_{\alpha\uparrow}+e^{-i\alpha\varphi/4}d_{\alpha\downarrow}\right)\right. \\
&\left.+c_{\alpha k-}^{\dagger}\left(-e^{i\alpha\varphi/4}d_{\alpha\uparrow}+e^{-i\alpha\varphi/4}d_{\alpha\downarrow}\right)\right\}+\text{H.c.}\ ,
\end{split}
\end{equation}
where $c_{\alpha k\pm}^{\dagger}$ ($c_{\alpha k\pm}$) creates (annihilates) an electron with momentum $k$ in lead $\alpha$ with majority/minority spin of lead $\alpha$. 
Again, we use the notation that whenever $\alpha$ is used as a variable rather than a coefficient, it takes the values $+1$ for $\text{L}$ and $-1$ for $\text{R}$. 
In this representation of the tunneling Hamiltonian it becomes apparent that 
the diagonal elements of the reduced density matrix couple to the off-diagonal ones with respect to the \textit{real} spin.
The reason for this coupling of diagonal and off-diagonal elements is that tracing out the ferromagnetic leads, we obtain dynamics for the reduced density matrix of the double dot which do not conserve spin.
Therefore, off-diagonal elements of the density matrix have to be taken into account here, and we consider the density-matrix elements $\boldsymbol{P}=\left(P_0,P_{\mathrm{b}\uparrow},P_{\mathrm{b}\downarrow},P_{\mathrm{a}\uparrow},P_{\mathrm{a}\downarrow},P_{\mathrm{b}\uparrow}^{\mathrm{b}\downarrow},P_{\mathrm{b}\downarrow}^{\mathrm{b}\uparrow},P_{\mathrm{a}\uparrow}^{\mathrm{a}\downarrow},P_{\mathrm{a}\downarrow}^{\mathrm{a}\uparrow}\right)$. The off-diagonal elements and the difference of the occupation number for different spins are contained in the vector
\begin{equation}
\vec{S}_{\eta}=\frac{1}{2}\left(\begin{array}{c}
      P_{\eta\downarrow}^{\eta\uparrow}+P_{\eta\uparrow}^{\eta\downarrow}\\
i(P_{\eta\downarrow}^{\eta\uparrow}-P_{\eta\uparrow}^{\eta\downarrow})\\
P_{\eta\uparrow}-P_{\eta\downarrow}\end{array}\right).
\end{equation}
for $\eta=\text{b},\text{a}$. The generalized master equation for the elements of the reduced density matrix reads
\begin{widetext}
\begin{subequations}
\begin{align}
\frac{ d}{ d t}P_{\eta}&=2\sum_{\alpha=\text{L},\text{R}}\left(\Gamma_{\alpha\eta}f^{+}\left(\epsilon_{\eta}\right)P_{0}-\vec{p}_{\alpha}\Gamma_{\alpha\eta}f^{-}\left(\epsilon_{\eta}\right)\vec{S}_{\eta}-\frac{1}{2}\Gamma_{\alpha\eta}f^{-}\left(\epsilon_{\eta}\right)P_{\eta}\right)\ , \\
\frac{ d}{ d t}\vec{S}_{\eta}&=\sum_{\alpha=\text{L},\text{R}}\left(\vec{p}_{\alpha}\Gamma_{\alpha\eta}f^{+}\left(\epsilon_{\eta}\right)P_{0}-\frac{1}{2}\vec{p}_{\alpha}\Gamma_{\alpha\eta}f^{-}\left(\epsilon_{\eta}\right)P_{\eta}-\Gamma_{\alpha\eta}f^{-}\left(\epsilon_{\eta}\right)\vec{S}_{\eta}\right)+\vec{B}_{\eta}\times\vec{S}_{\eta}\label{eq_spin}\ .
\end{align}
\end{subequations}
\end{widetext}
Again $P_0$ is obtained via the probability conservation $P_0^{(i)}+P_\mathrm{b}^{(i)}+P_\mathrm{a}^{(i)}=1$ and $P_0^{(a)}+P_\mathrm{b}^{(a)}+P_\mathrm{a}^{(a)}=0$. The last line in Eq.~(\ref{eq_spin}) represents a Bloch-type equation for the spin expectation value, which is affected by relaxation and accumulation of spin and by a rotation due to an effective magnetic field which reads
\begin{equation}
\vec{B}_{\eta}=\frac{1}{\pi}\sum_{\alpha=\text{L},\text{R}}\vec{p}_{\alpha}\Gamma_{\alpha\eta}\phi\left(\epsilon_{\eta}\right). 
\end{equation}
The function $\phi$ is defined by Eq.~(\ref{eq_phi}). This effective magnetic field is induced by coherent transitions between spin states and is different for the bonding and the antibonding channels.
Due to the absence of a bias voltage, we find that the \textit{instantaneous} density-matrix elements do not differ from the previous case without magnetization. In particular,  we find that the instantaneous spin expectation value is zero, $\vec{S}_{\eta}^{\left(i\right)}=0$. The \textit{adiabatic} correction to the spin is different from zero; nevertheless, in the absence of bias, the adiabatic spin vector $\vec{S}_{\eta}^{\left(a,-1\right)}$ turns out to be parallel to the effective magnetic field,~\cite{Splettstoesser08a} hence no precession takes place and the effective magnetic field does not influence the transport dynamics. 

We calculate the pumping current through the double-dot system in the presence of arbitrarily polarized ferromagnetic leads and find
\begin{equation}\label{eq_IL_ferro}
I_{\text{L}}\left(t\right)=e\sum_{\eta=\mathrm{b,a}}\frac{\Gamma_{\text{L}\eta}}{\Gamma_{\eta}}\left\{ 1+\Gamma_{\text{R}\eta}\Gamma\frac{\left(\vec{p}_{\text{R}}-\vec{p}_{\text{L}}\right)\vec{\pi}_{\eta}}{\Gamma_{\eta}^{2}-\Gamma^{2}\vec{\pi}_{\eta}^{2}}\right\} \frac{ d}{ d t}P_{\eta}^{\left(i\right)}\ ,
\end{equation}
with the definition of
\begin{equation}
\vec{\pi}_{\eta}=\frac{\Gamma_{\text{L}\eta}}{\Gamma}\vec{p}_{\text{L}}+\frac{\Gamma_{\text{R}\eta}}{\Gamma}\vec{p}_{\text{R}}.
\end{equation}
The sign of the correction term to the pumping current with respect to the nonmagnetic case [second term in Eq.~(\ref{eq_IL_ferro})] differs depending on the difference in the lead polarizations.
This means that the pumping current can be reduced in presence of differently polarized ferromagnetic leads, called the spin-valve effect,~\cite{Julliere75,Slonczewski89} or enhanced, showing an inverted spin-valve effect. Such an observation was  made for the pumping current through a single quantum dot attached to ferromagnetic leads.~\cite{Splettstoesser06}
In this case of a single dot, the spin-valve effect for the \textit{pumped charge} can only be inverted, when inducing a strong spatial asymmetry regarding the tunnel coupling. Considering a static transport bias, an inverted spin-valve effect has, for example, been found for a single quantum dot due to a spin dependence of interfacial phase shifts~\cite{Cottet06} as well as for a double quantum dot due to charge fluctuations in the nonlinear bias regime.~\cite{Trocha09}\\
Qualitatively very different features for the spin-valve effect are found when pumping through a double dot. We show in the following that this is due to effective spin-dependent asymmetries in the coupling to the different hybrid states.\\
For the pumped charge per area of the pumping cycle (in bilinear response) we find
\begin{equation}
\frac{N^\mathrm{p}}{A}=\sum_{\eta=\mathrm{b,a}}\frac{ d}{ d\bar{\epsilon}}\left(\bar{\Gamma}_{\text{L}\eta}\frac{\bar{\Gamma}_{\eta}-\Gamma\vec{p}_{\text{L}}\vec{\bar{\pi}}_{\eta}}{\bar{\Gamma}_{\eta}^{2}-\Gamma^{2}\vec{\bar{\pi}}_{\eta}^{2}}\right)\frac{ d}{ d \bar{E}}\bar{P}_{\eta}^{\left(i\right)}.
\end{equation}
We use the abbreviation $N^\mathrm{p}$ to indicate the pumped charge in presence of spin polarized leads to be contrasted with the pumped charge $N^0=N(\varphi=0)=N(p_\mathrm{L}=p_\mathrm{R}=0)$, in the presence of normal conducting leads [see Eq.~(\ref{eq_N_strong_Delta})]. 
According to Sec.~\ref{sec_strong_coupling} we separate the two contributions corresponding to pumping through the bonding and antibonding states, contributing with opposite signs to the pumped charge, $N_{\mathrm{b}}^\mathrm{p}/A>0$ and $N_{\mathrm{a}}^\mathrm{p}/A<0$.
The total pumped charge in the presence of polarized leads with respect to the pumped charge for vanishing polarization, $N^\mathrm{p}/N^0$, depends on the strength of the polarization of the two leads and on their respective angle $\varphi$.

\begin{figure}
\centering
\includegraphics[scale=1]{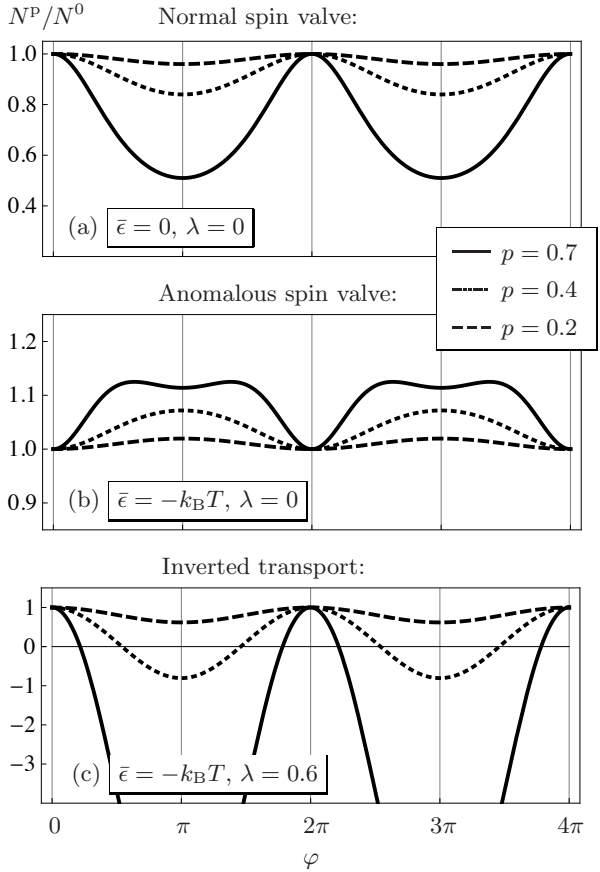}
\caption{Plot of the ratio of the pumped charge for polarized leads to the pumped charge for non-magnetic contacts $N^\mathrm{p}/N^0$, as a function of the magnetization angle $\varphi$ for increasing polarization strength $p=p_{\text{L}}=p_{\text{R}}=\left\{0.2,0.4,0.7\right\}$. (a) Plot for the symmetric case $\bar{\epsilon}=0$, $\lambda=0$.
(b) Plot for the case of $\bar{\epsilon}=-k_\mathrm{B}T$, $\lambda=0$.
(c) Plot for the strongly asymmetric case $\bar{\epsilon}=-k_\mathrm{B}T$, $\lambda=0.6$. For all three plots we take $\Delta=k_\mathrm{B}T$, $k_\mathrm{B}T=2\Gamma$.}
\label{fig_Ratio_Np_N0}
\end{figure}

The quantity $N^\mathrm{p}/N^0$ is independent of $\bar{E}$; it is shown for different parameter regimes in Fig.~\ref{fig_Ratio_Np_N0}. As a function of the angle $\varphi$, the relative pumped charge $N^\mathrm{p}/N^0$ is $2\pi$ periodic, and it equals one if $\varphi=0$. The stronger the polarization of the leads, the more the functional dependence on $\varphi$ deviates from a cosinelike behavior.~\cite{Slonczewski89}\\
Three different parameter regimes are distinguished in panels (a)-(c), where we choose $p_\mathrm{L}=p_\mathrm{R}=p$. The upper plot, Fig.~\ref{fig_Ratio_Np_N0}(a) shows results for a symmetric double dot, $\lambda=0$ and $\epsilon=0$, where the usual spin-valve effect is found. The spin-valve effect is most dominant when the left and right polarization vectors are in antiparallel alignment, $\varphi=\pi$.
When the double dot is made asymmetric, the spin-valve effect can be inverted, $N^\mathrm{p}>N^0$, ergo we find an anomalous spin-valve effect, see Fig.~\ref{fig_Ratio_Np_N0}(b). In this case, the deviation from the cosine behavior goes so far that for large polarizations, the maximum effect can even be found at angles away from $\pi$. This implies that in the limit $p\rightarrow 1$, the relative pumped charge equals one for $\varphi=0$ and zero elsewhere. The most surprising result is that even the transport direction can be inverted as compared to the normal case, leading to $N^\mathrm{p}/N^0<0$, see Fig.~\ref{fig_Ratio_Np_N0}(c). Also here, there is a  shift of the maximum effect away from $\varphi=\pi$ as in Fig.~\ref{fig_Ratio_Np_N0}(b), which becomes apparent for large polarizations and which is not shown in this figure.
\subsection{TMR of the pumped charge}
\begin{figure}
\centering
\includegraphics[scale=1]{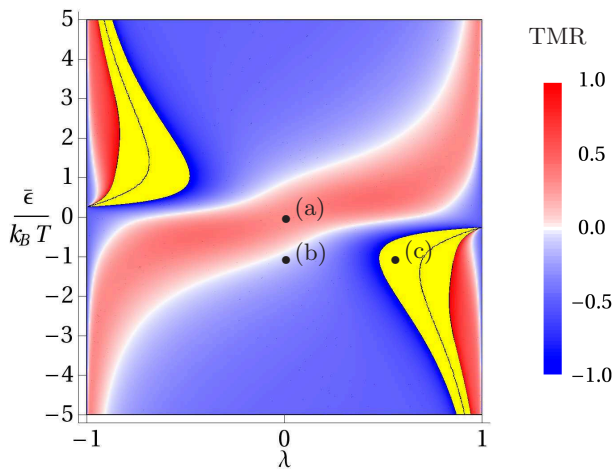}
\caption{(Color online.) Overview map of the tunneling magnetoresistance of the pumping system for $p=0.7$, $\varphi=\pi$ and $\Delta=k_\mathrm{B}T$; the TMR is independent of $\bar{E}$. Red represents the spin-valve effect, blue its inversion, and the yellow area denotes the region of inverted charge transport $\left|\text{TMR}\right|>1$. The black contours separating two yellow areas, represent the divergence of $\text{TMR}$ where $N^\mathrm{p}=-N^0$. Any contour separating blue (red) and yellow represents $N^0=0$ ($N^\mathrm{p}=0$). The black dots indicate the symmetry configurations chosen in Fig.~\ref{fig_Ratio_Np_N0}(a), (b) and (c).} 
\label{fig_TMR}
\end{figure}

To classify these three cases in a more convenient way we define in analogy to the static case~\cite{Julliere75,Trocha09} a  tunneling magnetoresistance (TMR) for pumping 
\begin{equation}
\text{TMR}=\frac{N^{0}-N^\mathrm{p}}{N^{0}+N^\mathrm{p}}.
\end{equation}
Values between $0$ and $1$ represent the case where transport is suppressed due to the spin-polarized leads, see the example in Fig.~\ref{fig_Ratio_Np_N0}(a). Values between $-1$ and $0$ on the other hand indicate enhanced transport (reversed spin-valve effect), see the example in Fig.~\ref{fig_Ratio_Np_N0}(b). In the more exotic case, where transport is reversed, the absolute value of the TMR is bigger than one and it diverges when $N^\mathrm{p}=-N^0$. The TMR is plotted as a function of the double-dot parameters in Fig.~\ref{fig_TMR}, where we show the case of antiparallel alignment of the magnetizations of the two leads, $\varphi=\pi$ and $p=p_\mathrm{L}=p_\mathrm{R}$. The three different regimes are represented by a color scale: the red area denotes the ordinary spin-valve effect ($0<\text{TMR}<1$), and in the blue area the spin-valve effect is reversed ($-1<\text{TMR}<0$). The parameter area in which the charge transport itself is inverted ($\left|\text{TMR}\right|>1$) is drawn in yellow, and the black line dividing the yellow area is indicating $N^\mathrm{p}=-N^0$. For comparison, the parameter choices for the three plots of Fig.~\ref{fig_Ratio_Np_N0} are indicated by the corresponding letters.

The inverted spin-valve effect, $-1<\mathrm{TMR}<0$, can be easily explained for this case of $\varphi=\pi$. Now $N_\uparrow$, the number of pumped electrons with spin up, relates, e.g., to the majority spin in the left lead and the minority spin in the right lead. The shift between $N_\uparrow$ and $N_\downarrow$, similar to what is shown in Fig.~\ref{fig_N_S_strong}, depends on the polarization strength, in particular, the shift is zero for $p=0$. This can lead to an increase or a decrease in the total pumped charge as a function of $\bar{\epsilon}$ and $\lambda$, implying that the total pumped charge in the presence of polarized leads, $N^\mathrm{p}$, can be larger than for normal conducting leads, $N^0$. This effect can  occur independently of the accessibility of the excited (antibonding) state, as it does not rely on a sign change in $N_\uparrow$ or $N_\downarrow$.

When $\left|\bar{\epsilon}_\mathrm{b}-\bar{\epsilon}_\mathrm{a}\right|\lesssim k_\mathrm{B}T$ parameter regimes exist where the transport changes direction, i.e., $N^\mathrm{p}$ and $N^0$ have different signs [as depicted in Fig.~\ref{fig_Ratio_Np_N0}(c)].
As we have stated before, the polarization changes the transport through the spin-dependent bonding and antibonding channels differently. Consequently the transport through the excited antibonding state becomes dominant at different parameter configurations, depending on the polarization of the leads. Therefore we can obtain different nodes for $N^0_{\text{b}}+N^0_{\text{a}}=0$ (contours separating yellow and blue areas in Fig.~\ref{fig_TMR}) and  $N^\mathrm{p}_{\text{b}}+N^\mathrm{p}_{\text{a}}=0$ (contours separating yellow and red areas in Fig.~\ref{fig_TMR}). In between these nodes, the charge transport is reversed. The regions of inverted transport can be enlarged by increasing polarization strengths $p$.  
 
We compare this with the tunneling magnetoresistance for a double-dot system subject to a static bias in the linear-response regime. All parameters are therefore time independent. We define the static tunneling magnetoresistance as
\begin{equation}\label{eq_TMR_stat}
\text{TMR}_\text{stat}=\frac{G^{0}-G^\mathrm{p}}{G^{0}+G^\mathrm{p}}\ ,
\end{equation}
via the linear conductance in the presence of normal leads, $G^0$,  and in the presence of spin-polarized leads, $G^\mathrm{p}$. Their explicit form is given in Appendix~\ref{app_stat_TMR}. One can show that for all possible parameters in the regime of strong interdot coupling
\begin{equation}
0<\text{TMR}_\text{stat}<1\ .
\end{equation}
This means that the spin-valve effect in the \textit{static linear-response regime} in lowest order in the tunnel coupling cannot be reversed. 

An inverted spin-valve effect was found even without time-dependent parameters for spin-depedent tunneling~\cite{Cottet06} and for nonlinear bias.~\cite{Trocha09} In contrast, the \textit{charge transport inversion}, which we find for the pumped charge, corresponding to $|\mathrm{TMR}_\mathrm{pump}|>1$,
is not expected  in a static situation, because it is induced by thermal transport through the excited state. In the static linear response, thermal transport through an excited state is always a minor correction to transport through the ground state. In contrast, we showed in this section that due to the time dependence of the system parameters one can realize certain configurations (tuned by the double dot's parameters $\bar{\epsilon}$ and $\lambda$) in which the contribution from transport through the ground state is suppressed in the time-averaged pumped charge.

We finally mention results for the TMR in the weak interdot coupling regime, which is not discussed in detail here. 
Also in this case the $\text{TMR}$  takes values in the whole range $[-\infty,\infty]$, which can be shown directly from a study of the N-DD-F  system in this regime: already if only one lead is polarized, we find a reversed spin-valve effect as well as reversed charge transport in the presence of polarized leads, due to polarization dependent quantum charge fluctuations.

A multiple quantum dot in a carbon nanotube in contact with ferromagnetic leads was recently realized experimentally~\cite{Feuillet10}  - a system which is at the basis of our study in Secs.~\ref{sec_spin_pump} and \ref{sec_ferro}. The application of time-dependent fields to such a setup would therefore allow to experimentally verify our results regarding spin-dependent transport.

\section{Conclusion}
We have investigated charge and spin pumping through a serially coupled double quantum dot in both the weak and strong interdot coupling regimes, contacted to either normal or ferromagnetic reservoirs. We take into account Coulomb interaction with the only restriction that double occupation of a single dot is excluded. When the two dots are weakly coupled, interaction-induced quantum charge fluctuations occur, resulting in a renormalized level difference. We show that this level renormalization can be directly measured as a node in the pumped charge. In the case of strong interdot coupling, we find that hybridized double-dot states contribute to the transport with opposite sign. If their level spacing is within thermal reach, the time-averaged transport of the excited channel can outmatch the ground-state one. This results in a sign change in the averaged pumped charge, uniquely found in the presence of time-dependent fields.\\
Including one ferromagnetic contact, both mechanisms, i.e., quantum charge fluctuations in the weak interdot coupling and thermal accessibility of the excited state in the strong interdot coupling, enable the possibility to pump spin in the absence of net charge transport.\\ 
We finally studied the spin-valve effect for non-collinear ferromagnetic contacts. Depending on the system's parameters, we find a normal as well as an anomalous spin-valve effect, and even an actual change in the transport direction can be observed, leading to a diverging tunneling magnetoresistance.

\acknowledgments
The authors would like to thank A. Cottet, M. Governale,  G. G\"untherodt, J. K\"onig, T. Kontos, F. Reckermann, and M. R. Wegewijs for helpful comments and valuable discussions. We acknowledge financial support by  the Ministry of Innovation NRW. 

\begin{appendix}
\section{Perturbation expansion for weak interdot coupling}\label{app_pt_weak}
The single-electron eigenstates of the isolated double-dot system are given by the bonding and the antibonding states, see Sec.~\ref{sec_strong_coupling}. If the level spacing between these two states is large with respect to the coupling to the leads  
$\Gamma$, off-diagonal elements of the reduced density matrix become fundamentally important only in second order in $\Gamma$ (Ref.~\onlinecite{Leijnse08}) but are suppressed in lowest order in the tunnel coupling, $\Gamma<k_\mathrm{B} T$. The present work is limited to the case of weak coupling, where a description up to first order in $\Gamma$ is justified. In Secs.~\ref{sec_weak_coupling} and~\ref{sec_spin_pump_weak}, we are interested in the regime where the splitting of the two levels is small and coherent oscillations between the related states play an important role already in first order in $\Gamma$. Therefore, the interplay of the internal dynamics and the dynamics due to the coupling to the leads is most interesting when all three parameters are of the same order, $\Delta\sim\epsilon\sim\Gamma$, and thus have to be considered on the same level of approximation.\cite{Wunsch05}\\
In this regime, when coupling between the left and the right dot is therefore small, $\Delta\lesssim \Gamma$, it is intuitive to describe the double-dot system by the states $|\mathrm{L}\sigma\rangle$ and $|\mathrm{R}\sigma\rangle$. These are the eigenstates of the \textit{fully decoupled} single dots, and the corresponding basis does not represent an exact eigenbasis of the isolated double-dot system.\\
The localized states differ from the hybrid bonding and antibonding states only when $\Delta$ is taken into account in at least first order. On the other hand  the Kernel $\boldsymbol{W}$, entering the last term of the generalized master equation, Eq.~(\ref{eq_master_general}), describes tunneling to the leads, and therefore starts in first order in $\Gamma$. When treating the parameters $\Delta$, $\epsilon$ and $\Gamma$ on the same footing, for the calculation of the Kernel, the eigenstates of the double-dot system are therefore taken into account in zeroth order in the small parameters and equal $|\mathrm{L}\sigma\rangle$ and $|\mathrm{R}\sigma\rangle$ within the limits of the approximation. This additionally justifies the description of the dynamics of the system in terms of the localized eigenstates,  $|\mathrm{L}\sigma\rangle$ and $|\mathrm{R}\sigma\rangle$.

\section{Static TMR}\label{app_stat_TMR}
We calculate the conductance for the static case [see Eq.~(\ref{eq_TMR_stat})] in the linear-response regime. We focus on strong interdot coupling $\Delta>\Gamma$ and infinite Coulomb interactions $U$ and $U'$ in order to compare to the results found in Sec.~\ref{sec_ferro}. The linear conductance in the presence of polarized leads, $G^\mathrm{p}$, is obtained by solving the stationary master equation, Eq.~(\ref{eq_master_stat}) for leads with different Fermi energies and by calculating the stationary current response from Eq.~(\ref{eq_current_general}).
We find for the conductance
\begin{align}\label{eq_G_lin}
G^\mathrm{p}=&2e^2\beta\sum_{\eta=\mathrm{b,a}}f^{+}\left(\epsilon_{\eta}\right)\frac{\Gamma_{\text{L}\eta}\Gamma_{\text{R}\eta}}{\Gamma_{\eta}}\left(1-\frac{\Gamma_{\text{L}\eta}\Gamma_{\text{R}\eta}}{\Gamma_{\eta}^{2}-\Gamma^{2}\vec{\pi}_{\eta}^{2}}\right. \nonumber\\
 &\left.\times\frac{C\Gamma_{\eta}^{2}\left[f^{-}\left(\epsilon_{\eta}\right)\right]^2+\left(\left(\vec{p}_{\text{L}}-\vec{p}_{\text{R}}\right)\vec{B}_{\eta}\right)^{2}}{\Gamma_{\eta}^{2}\left[f^{-}\left(\epsilon_{\eta}\right)\right]^2+\vec{B}_{\eta}^{2}}\right)P_{0}\ ,
\end{align}
with
\begin{equation}
C=\left(\vec{p}_{\text{L}}-\vec{p}_{\text{R}}\right)^{2}+\left(\vec{p}_{\text{L}}\vec{p}_{\text{R}}\right)^{2}-p_{\text{L}}^{2}p_{\text{R}}^{2}\ ,
\end{equation}
and the static probability of having an empty dot in lowest order in tunneling $P_{0}=(1+2e^{-\beta\epsilon_\mathrm{b}}+2e^{-\beta\epsilon_\mathrm{a}})^{-1}$. The linear conductance for unpolarized leads $G^{0}$ is found by setting the polarizations to zero, making the correction term in the parentheses vanish. 
We insert the linear conductances found from Eq.~(\ref{eq_G_lin}) into Eq.~(\ref{eq_TMR_stat}) in order to find $\mathrm{TMR}_\mathrm{stat}$. 
The static tunneling magnetoresistance depends on $\bar{E}$ in contrast to the pumping $\mathrm{TMR}$. Furthermore, due to the applied bias, the effective magnetic field enters in the static case, which it does not in the case of pumping for vanishing bias. 
In contrast to the pumping case, $\mathrm{TMR}_\mathrm{stat}$ takes always values within in the interval $[0,1]$.
\end{appendix}

\bibliographystyle{apsrev4-1}
\bibliography{cite}

\end{document}